\documentclass[pra,twocolumn,eqsecnum,superscriptaddress,floatfix,showpacs]{revtex4}

\usepackage{graphicx}% Include figure files
\usepackage{dcolumn}% Align table columns on decimal point
\usepackage{bm}% bold math
\usepackage[usenames]{color}
\newcommand{\ket}[1]{|#1\rangle}
\newcommand{\erf}[1]{\textrm{erf}\left(#1\right)}
\newcommand{\e}{+1}
\newcommand{\g}{-1}
\newcommand{\rec}{{\psi}}

\newcommand{\nn}{\nonumber}
\newcommand{\nl}{\nn \\ &&}

\newcommand{\eqrf}[1]{Eq.~(\ref{#1})}
\newcommand{\eqrfs}[2]{Eqs.~(\ref{#1}) and (\ref{#2})}

\begin{document}
\title{Protocols for optimal readout of qubits using a continuous \\quantum nondemolition measurement}
\author{Jay Gambetta}
    \affiliation{Departments of Applied Physics and Physics, Yale
    University, New Haven, CT 06520}
\author{W. A. Braff}
     \affiliation{Departments of Applied Physics and Physics, Yale
    University, New Haven, CT 06520}
\author{A. Wallraff}
    \affiliation{Departments of Applied Physics and Physics, Yale
    University, New Haven, CT 06520}
    \affiliation{Department of Physics, ETH Zurich, CH-8093 Zurich, Switzerland}
\author{S. M. Girvin}
    \affiliation{Departments of Applied Physics and Physics, Yale
    University, New Haven, CT 06520}
\author{R. J. Schoelkopf}
    \affiliation{Departments of Applied Physics and Physics, Yale
    University,  New Haven, CT 06520}
\date{\today}

\begin{abstract}
We study how the spontaneous relaxation of a qubit affects a
continuous quantum non-demolition  measurement of the initial state of the qubit.
Given some noisy measurement record $\Psi$, we seek an
estimate of whether the qubit was initially in the ground or excited
state.  We investigate four different measurement protocols, three of
which use a linear filter (with different weighting factors) and a
fourth which uses a full non-linear filter that gives the
theoretically optimal estimate of the initial state of the qubit. We
find that relaxation of the qubit at rate $1/T_1$ strongly
influences the fidelity of any measurement protocol. To avoid errors due to this decay, the measurement
must be completed in a time that decrease linearly with the desired fidelity while maintaining an adequate signal to noise ratio.
We find that for the non-linear filter the predicted fidelity, as expected, is always
better than the linear filters and that the fidelity is a monotone
increasing function of the measurement time.
For example,  to achieve a fidelity of 90\%, the box car linear filter
requires a signal to noise ratio of $\sim 30$  in a time $T_1$ whereas the non-linear
filter only requires a signal to noise ratio of $\sim 18$.
\end{abstract}


\maketitle

\section{\label{sec:intro}Introduction}

{In this paper we consider the following problem. Given a qubit initially
either in the ground or excited state with finite lifetime $T_1$, how can we
best make use of a continuous-in-time noisy quantum non-demolition (QND) measurement to optimally estimate
the \emph{initial} state (ground or excited) of the qubit? A number of authors have previously
considered the related problem of optimal estimation of the \emph{present} state of the qubit based on the past and
current measurement record \cite{Car93,makhlin00, korotkov01, makhlin01, korotkov01b,goan:2001}, 
but, to the best of our knowledge, 
the problem we consider here has not been previously studied.}

QND measurements play a central role in the
theory and practical implementation of quantum measurements
\cite{AsherPeres}. In a QND measurement, the interaction term in the
Hamiltonian coupling the system to the measuring apparatus commutes
with the quantity being measured, so that this quantity is a constant
of the motion. This does \emph{not} imply that the quantum state of
the system is totally unaffected, but it does imply that the
measurement is {\em repeatable}.  For example, a Stern-Gerlach
measurement of $\hat\sigma_z$ for a spin 1/2 particle initially prepared
in an eigenstate of $\hat\sigma_x$ will randomly yield the results $+1$
and $-1$ with equal probability.  However, all subsequent
measurements of $\hat\sigma_z$ will yield exactly the same result as the
initial measurement.

The fact that QND measurements are repeatable is of fundamental
practical importance in overcoming  detector inefficiencies. A
prototypical example is the electron-shelving technique
\cite{nagourney86,sauter86} used to measure trapped ions.  A related
technique is used in present implementations of ion-trap based
quantum computation.  Here the (extremely long-lived) hyperfine
state of an ion is read out via state-dependent optical
fluorescence. With properly chosen circular polarization of the
exciting laser, only one hyperfine state fluoresces and the
transition is cycling; that is, after fluorescence the ion almost
always returns to the same state it was in prior to absorbing the
exciting photon.
 Hence the measurement is QND.  Typical experimental parameters
 \cite{WinelandNISTJRESExpMethods}
 allow the cycling transition to produce $N\sim 10^6$ fluorescence
 photons.  Given the photomultiplier quantum efficiency and typically small solid
 angle coverage, only a very small number $\bar n_{\rm d}$ will be detected on
 average.  The probability of getting zero detections (ignoring dark
 counts for simplicity) and hence misidentifying the hyperfine state
 is $P(0)=e^{-\bar n_{\rm d}}$.  Even for a
 very poor overall detection efficiency of only $10^{-5}$, we still have $\bar n_{\rm
 d}=10$ and nearly perfect fidelity $F=1-P(0)\sim 0.999955$.
  It is important to note that the total time available for measurement is
not limited by the phase coherence
 time ($T_2$) of the qubit or by the
 measurement-induced dephasing~\cite{korotkov01,makhlin01,schusterdephasing,GamBlaSch06},
 but rather only by the rate at which the qubit makes real transitions between measurement
 ($\hat\sigma_z$)
 eigenstates.  In a perfect QND measurement there is no measurement-induced state
 mixing \cite{makhlin01} and the relaxation rate $1/T_1$ is unaffected
 by the measurement process.

The ability to read out a qubit with high fidelity is of central
importance to the successful construction of a quantum computer
\cite{nielsen00}. In order to successfully measure a qubit, its
quantum state must be mapped into a piece of classical information
by measuring the relative occupation of its two states with the
highest possible fidelity. Possible qubit
implementations include superconducting circuits,
silicon based electron and nuclear spins, and trapped ions,
among
others~\cite{kane98,vanderwal00,vrijen00,vion02,chiorescu03,astafiev04,langer05,blais04,wallraffNature,wallraff05}.
In order for qubits prepared in different states to be distinguishable,
the measurement must be completed before
the excited qubits decay\cite{divincenzo99,makhlin01}. Many atomic qubits have
sufficiently long lifetimes so
that relaxation is not a major concern~\cite{langer05,kane98,vrijen00},
but most solid state qubits have
lifetimes on the order of microseconds or less, and spontaneous relaxation plays a
significant role in the
measurement. The qubit relaxation affects different measurement schemes differently,
but in all cases, it can limit the maximum fidelity.

%%{\red Steve i really feel we said all the points we wanted to}
Although the behavior of a qubit during continuous measurement has
been studied using Monte Carlo simulations~\cite{makhlin00, korotkov01, makhlin01, korotkov01b,goan:2001}, no attempt
has been made to derive an analytical expression for the probability
distribution of the \emph{initial} state or to study how spontaneous emission
impacts the measurement fidelity. This is at least partially because
there exist very few high fidelity continuous QND measurements, and
even fewer that operate in a regime where qubit relaxation is a
limiting factor. With the application of low temperature amplifiers,
high fidelity (though not necessarily weak continuous)
measurements of superconducting qubits are now becoming
feasible~\cite{astafiev04,wallraff05,loss03,cooper04,bertet04,martinis02,martinis_katz06,siddiqi04,siddiqi05,siddiqi06}.
These measurements have found asymmetries
between the probability distributions for the integrated signal
corresponding to the ground and excited states, but could not
accurately predict them.

Here we find that when we use a measurement protocol that only records the integrated signal,
the qubit
relaxation induces asymmetry in the probability distributions, and that with a
sufficiently precise detector,
the distributions become distinctly non-Gaussian. Unlike measurement of a perfect (i.e., non-decaying) qubit,
where fidelity is always improved by a longer measurement, we show that there is some optimal measurement
fidelity  that depends on the signal to noise ratio (SNR) of the detector and the filter used. The
first filter we consider is the linear box car filter and optimize over the integrated time $t_{\rm f}$.
Choosing a longer or shorter integration time will lower the fidelity of the
measurement. Next we show that by choosing a filter that gives exponentially less importance to results at later
times slightly increases the fidelity.
We then numerically find the optimal linear filter and
compare these linear filters to a
non-linear filter that yields the theoretically optimal estimate of the initial state of the qubit
given some measurement record $\Psi$. We find that we can reach the same
fidelity as the linear filters at a substantially lower SNR. Furthermore, due to the nature of the updating
protocol, the fidelity is a non-decreasing function of the measurement time.
In summary, in this paper we determine the optimal
measurement fidelity given four measurement protocols for
continuous measurement experiments currently being performed, and
also provides a guideline for the necessary detector signal to noise
ratio in order to reach a particular desired fidelity in future
experiments.

%Jay addition
There are two major ways of measuring qubits. The first method is a
latching measurement, for example by having the qubit state modify
the switching current (or state) of an adjacent Josephson junction
~\cite{martinis02,martinis_katz06}
or the bifurcation point of the non-linear Josephson plasma oscillation \cite{siddiqi04,siddiqi05,siddiqi06}. In
such latching measurements, the qubit is measured very quickly with very high signal to noise, and after only a
short waiting time~\cite{martinis02,vion02,vanderwal00,siddiqi06}.  Some versions of such strong measurements
can in principle be QND \cite{siddiqi06}.
The second method is to perform a sequence of repeated or weak
continuous quantum non-demolition (QND) measurements, which each
leave the populations of the qubit unchanged.
 Several recent experiments with solid-state
 qubits \cite{wallraff05,duty04,lehnert04}, have used continuous QND measurement schemes in
 which the qubit lifetime imposes the main limitation on the measurement fidelity, for which
 the analysis of this paper should apply.

It is not just continuous measurements that are affected by qubit
relaxation. For example, consider an idealized latching measurement
scheme where a qubit is prepared in an eigenstate, but there is some
finite arming time $t_{\rm{arm}}$ before a perfect measurement is
made. In this case, a qubit prepared in the ground state will always
be measured correctly, but a qubit prepared in the excited state may
have decayed during $t_{\rm{arm}}$ and be misidentified. Thus if
$t_{\textrm{arm}}$ is not infinitesimally small, the qubit lifetime
places a limit on fidelity even with a perfect detector.

\section{\label{sec:classical}Qubit with infinite lifetime}

It is worthwhile to first consider the case when the qubit cannot relax from the state it is
initially prepared in, so it is ``fixed'' for all time. This allows us to formalize our
intuitive understanding of a general continuous measurement and to give us a result against which we can compare
the finite lifetime case. We consider the measurement of a qubit with two states $\ket{+}$ (excited) and
$\ket{-}$ (ground) and assume that the measurement result is given by the actual value of the qubit state plus
Gaussian noise.  This assumption is justified for example in the current circuit quantum electrodynamic experiments \cite{wallraffNature,schusterdephasing,wallraff05,GamBlaSch06} in which a cavity
is dispersively coupled to the qubit (no energy is
exchanged between the cavity and the qubit).  A homodyne measurement on the cavity output will reveal the cavity state (which is proportional to $\hat\sigma_z$) plus Gaussian noise \cite{blais04,GamBlaSch06}. This Gaussian noise will be at least the photon shot noise but in present experiments it is dominated by the following amplifier. In other words, 
the measurement is faithful and given that the system is in state $i =\pm 1$ our detector for a time
interval $d\tau$ outputs $\rec(\tau)$ with statistics
\begin{equation}\label{psigz}
P(\rec|i) = \sqrt{\frac{d \tau ~ {\rm SNR}}{2 \pi }} \exp[-(\rec -
i)^2 d \tau ~ {\rm SNR}/2 ].
\end{equation}   For convenience we have also introduced a dimensionless time
$\tau = t/T_1$ where $T_1$ is an arbitrary but finite number, that will become the relaxation lifetime when we
treat the finite lifetime case.  Here  SNR is the ratio of integrated signal power to noise power.  It is linear
in the integration time and we will adopt the convention of specifying the SNR as that achieved after
integrating for time $T_1$.
 %and is defined
%to be the signal power divided by the noise power, so SNR is linear in the integration time.

From this distribution we can  write $\psi(\tau)$ in terms of the Wiener increment $d W(\tau)$ \cite{Gar85} as
\begin{equation}\label{psific}
\psi (\tau) d \tau =   i_\pm (\tau)d \tau+ \sqrt{{\rm SNR}^{-1}} d
W(\tau) .
\end{equation}  Here we have introduced the subscript $\pm$ to indicate a possible realization of
the dynamics of the qubit given the initial condition $\pm 1$. For this case the qubit can be
initialized in either state, but because it has an infinite
lifetime, it is fixed in whatever state it starts in for the
duration of the measurement.  That is, $i_\pm(\tau) = \pm 1$.

We define our measurement signal $s$ as the output of the
detector
integrated over time $\tau_{\rm{f}}$
\begin{equation} \label{eq:flat}
s = \int_0^{\tau_{\rm{f}}}d\tau\, \psi(\tau).
\end{equation}
Formally we are restricting ourselves here to a simple box car linear filter
which uniformly weights the measurement record $\psi(\tau)$ in the
interval $0<\tau<\tau_{\rm{f}}$.  Using \eqrf{psific} it is simple
to carry out the above integral and rewrite the measurement signal as
$s_\pm(\tau_{\rm{f}})=\pm \tau_{\rm{f}} + X_\mathrm{G} [0, \sigma^2]$
where $X_\mathrm{G} [0, \sigma^2]$ is a Gaussian random variable of
mean 0 and standard deviation $\sigma= \sqrt{\tau_{\rm f}/{\rm
SNR}}$. The signals then follow the familiar Gaussian distributions
\begin{eqnarray}
    \label{eq:classical}
    P^{\rm{fixed}}_{\pm}(s) = \frac{1}{\sigma
    \sqrt{2\pi}}
    e^{-(s\mp\tau_{\rm{f}})^2/(2\sigma ^2)}.
\end{eqnarray}
 Because these
distributions are symmetric about $s=0$, the most obvious analysis
is to set a signal threshold $\nu_{\rm{th}}=0$ and call every
measurement with $s>\nu_{\rm{th}}$ a ($+1$) state, and every measurement
with $s<\nu_{\rm{th}}$ a ($-1$) state. Calculation of fidelity in this
case is accomplished using the definition

\begin{eqnarray}
    \label{eq:fidelity}
    F&\equiv&1-\int_{-\infty}^{\nu_{\rm{th}}}ds\, P_+(s) -
    \int_{\nu_{\rm{th}}}^{\infty}ds\, P_-(s).
\end{eqnarray}
For the case of infinite qubit lifetime we have the simple result
\begin{equation}
    F={\rm erf}\left({\sqrt{\frac{\tau_{\rm{f}} \rm{SNR}}{2}}}\right).
\end{equation}

A fidelity of zero corresponds to a completely random measurement
that extracts no information, a fidelity of one corresponds to a
perfect faithful measurement, and in between the measurement conveys varying
degrees of certainty. As $\tau_{\rm{f}}$ becomes large, Eq.~(\ref{eq:fidelity})
predicts that the fidelity rapidly
approaches unity. Higher SNR serves to speed up the convergence, but
as long as SNR is non zero, any desired fidelity is attainable
simply by measuring the qubit for long enough. In
Table~\ref{tab:fidelity}, the required $\rm{SNR}_{\rm{fixed}}$ is
listed in order to achieve a given fidelity within $T_1$.
Note the same results for the fidelity would be obtained if we used the optimal non-linear filter of Sec.
\ref{sec:NL}. That is, for a qubit with infinite lifetime the simple box car filter is
equivalent to the optimal non-linear filter, it is only when we include relaxation this is not the case. This will
be discussed in detail in the next four sections.

\section{\label{sec:Flat}Box car linear filter}

\subsection{\label{sec:quantum}Probability distributions for QND measurement}

\begin{figure}
      \includegraphics*[width=0.45\textwidth]{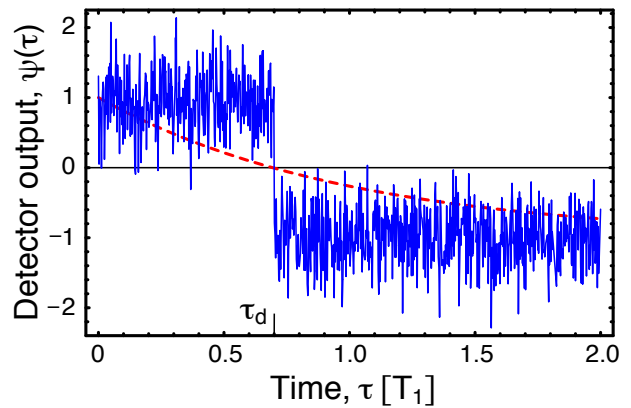}
    \caption{\label{fig:spontemission}(Color online) {\small Example detector output as a function of time
    for a continuous QND measurement with SNR = 570, ($d\tau = 1\times 10^{-4}$) which allows for
    a measurement fidelity of 99\%.
    The dashed red line corresponds to a large ensemble average of
    qubit outputs, and decays exponentially with characteristic time
    $T_1$. The solid blue line corresponds to the measurement record for a particular single
    shot.
    At this value of SNR, the jump in output is clearly visible and occurs at $\tau_{\rm{d}}$.}
    }
\end{figure}

 Here we consider the same measurement protocol as in Sec. \ref {sec:classical},
 however unlike a qubit with infinite lifetime, where both states behave
similarly,
a qubit with a finite lifetime
has a fundamental asymmetry in how the excited and ground states behave.  We assume (in our model) that the
excited state decays at rate $1/T_1$ but the transition rate upward out of the ground state is zero. A qubit
prepared in the ground state will never experience any excitations, so it can be treated much the same as the
fixed qubit discussed above and hence $P_-(s)=P_-^{\rm{fixed}}(s)$. By contrast, an initially excited qubit will
produce an ensemble averaged output which will decay exponentially with characteristic dimensionless time
$\tau_1\equiv 1$.
Although most qubits dephase in some shorter dimensionless time
$\tau_2$, our only concern here is the relative population of the
two qubit states, so we are not limited by decoherence of the qubit.
For a single qubit, this translates into a single, abrupt relaxation
of the qubit at some dimensionless time $\tau_{\rm{d}}$ that is
exponentially distributed with mean dimensionless time 1,
$P(\tau_{\rm{d}})=\exp(-\tau_{\rm{d}})$. That is, if the
qubit was initially in the excited state then $i_\pm(\tau)$ would obey
\begin{equation} \label{if_excited} i_\pm (\tau)= \theta(\tau_
{\rm{d}} - \tau) - \theta(\tau-\tau_ {\rm{d}}) .
\end{equation} Thus given a possible realization $i_\pm(\tau)$,  we can generate a typical record an experimentalist
would measure by using \eqrf{psific}. A typical trajectory for
$\psi(\tau)$ is shown in Fig.~\ref{fig:spontemission} for a ${\rm
SNR}$ of $570$. Here we see, at this value of SNR, the jump in output
is clearly visible and occurs at $\tau_{\rm{d}}\neq 1$ ($t_{\rm{d}}\neq T_1$).

In the case where the qubit happens to decay early, $\tau_{\rm{d}}\ll 1$, the signal $s$ from a qubit
initially prepared in the excited state would be almost indistinguishable from that of a qubit initially
prepared in the ground state, and even if the measurement apparatus were nearly perfect, almost no information
could be extracted.
The probability distribution $P_+(s)$ for the measurement signal of
an initially excited qubit can be determined analytically with a
simple derivation. This will allow a more quantitative discussion of
fidelity, and will eventually allow for optimization of integration
time, $\tau_{\rm f}$. The critical difference between the infinite lifetime system
described in Sec.~\ref{sec:classical} and an actual QND
measurement is that $s$ is now a function not only of the
dimensionless integration time $\tau_{\rm{f}}$, but also the
exponentially distributed random relaxation time
$\tau_{\rm{d}}$.

If the qubit is initially in the excited state, then from Eqs.
(\ref{psific}), (\ref{eq:flat}),  and (\ref{if_excited}) a possible
realization of $s_\pm$ will be
\begin{eqnarray}
s_\pm &=&  \tau_{\rm{f}} \theta(\tau_ {\rm{d}} -\tau_{\rm{f}})
	+ (2 \tau_ {\rm{d}} -\tau_{\rm{f}}) \theta(\tau_{\rm{f}} -
			\tau_ {\rm{d}}) \nl+ X_\mathrm{G} [0, {\sigma^2}].
	\end{eqnarray}   That is, the probability distribution for $s$ given a realization with a decay at
	time $\tau_{\rm d}$ is
	\begin{eqnarray}
	P_{+}(s|\tau_{\rm d})  &=& \frac{1}{\sqrt{2 \pi }\sigma} \exp[-(s- \tau_{\rm
			f} )^2/ 2\sigma^2] \theta(\tau_ {\rm{d}} -\tau_{\rm f})
	\nl+\frac{1}{\sqrt{2 \pi}\sigma} \exp\{-[s- (2\tau_ {\rm{d}} -
			\tau_{\rm f}) ]^2/ 2\sigma^2\} \nl\times\theta(\tau_{\rm f} -
				\tau_ {\rm{d}}),
			\end{eqnarray}  and from this one can easily obtain the probability distributions for $s$
			by averaging over all possible realizations (decay times). Doing this gives
			\begin{eqnarray}
			&P_+(s) = \frac{1}{\sqrt{2\pi} \sigma} e^{-(s-\tau_{\rm{f}})^2/(2\sigma^2)-\tau_{\rm{f}}}
			+\frac{1}{4}e^{-(s+\tau_{\rm{f}})/2} \nonumber& \\
				& \times e^{\sigma^2/8}\{\erf{\frac{\sigma^2-2(s-\tau_{\rm{f}})}{2\sqrt{2}\sigma}}-
				\erf{\frac{\sigma^2-2(s+\tau_{\rm{f}})}{2\sqrt{2}\sigma}}\}&
				\label{eq:probdist}
				\end{eqnarray}

				Although it does not figure directly into this analysis, it is not
				difficult to expand this treatment to consider a measurement with
				finite ``demolition" that stimulates both excitation and relaxation
				of the qubit. In this case, both $P_+$ and $P_-$ will be
				non-Gaussian, because the qubit may excite and relax several times
				during the measurement interval. To calculate the distributions,
				all we need to do is extend the possible realization to include multiple relaxations and excitations
				keeping in mind that the relaxation and excitation times
				are not independent variables: $\tau_{\rm{d}_n}$ must occur
				before $\tau_{\rm{d}_{n+1}}$. For stronger or less ideal
				measurements, it may become necessary to include these extra terms,
				but for now the demolition is taken to be small compared to the
				spontaneous relaxation and can be ignored.

				\begin{figure}
				\includegraphics[scale=.40]{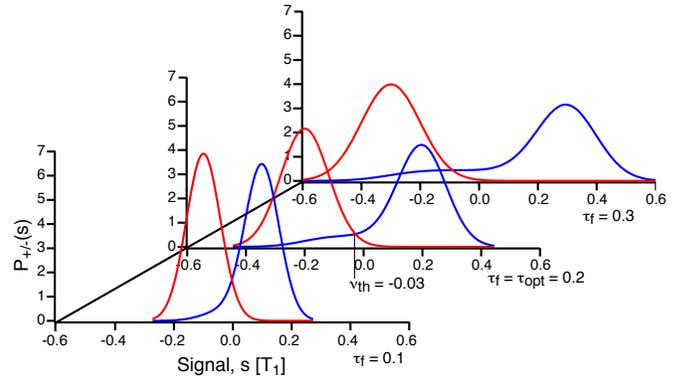}
	\caption{\label{fig:highNevo} (Color online) {\small Evolution in time of probability distributions for
		QND measurement of qubits initialized in the ground (blue)
			and excited (red) states for the box car linear filter and a $\rm{SNR}=30$. For
			short $\tau_{\rm{f}}$, the distributions are almost symmetrical
			because an excited qubit has probably not decayed yet. As
			$\tau_{\rm{f}}$ gets longer, the qubit is much more likely to
			have decayed, and the mean of $P_+(s)$ begins to drift back
			towards that of $P_-(s)$. At some optimal point in between,
					$\tau_{\rm{f}}=\tau_{\rm{opt}}$,
					and if the threshold $\nu_{\rm{th}}$ is chosen so that $P_+(\nu_{\rm{th}})=P_-(\nu_{\rm{th}})$, the
						misidentified tails are minimized, and fidelity is maximized.}
	}
\end{figure}

The Gaussian first term in Eq.~(\ref{eq:probdist}) is dominant for
$\tau_{\rm{f}}\ll1$, and $P_+(s)$ is nearly symmetrical to
$P_-(s)$. For these fast measurements, there is little chance that
the qubit decays during the measurement, and the distributions will
be very similar to those of fixed state bits. At some point, the
probability that the qubit has relaxed will be large enough that it
significantly affects the distributions. At this point, predictions
based on the assumption of no relaxation are no longer valid, and
the non-Gaussian term in Eq.~(\ref{eq:probdist}) becomes very
important. This effect can be seen in the non-Gaussian tail of
$P_+(s)$ in the second two time cuts in Fig.~\ref{fig:highNevo}.

A strongly non-Gaussian $P_+$ distribution is not required for the
fidelity to exhibit a maximum at some finite measurement time. Even
though for experiments with $\mathrm{SNR}\approx 1$, the non-Gaussian tail
of $P_+$ is not very prominent, the asymmetry between the
distributions in both height and width is quite clear.
For long enough
dimensionless integration times, a qubit initially in the excited
state will have probably relaxed relatively early in the
measurement, so its mean should be very similar to that of a qubit
initially in the ground state, the distributions will be almost
identical, and all resolving power will be lost.

\subsection{\label{sec:timefidelity}Optimal box car filter}

The behavior of fidelity as a function of integration time for a QND
qubit measurement is very different from a measurement of a fixed
state qubit that never relaxes. Recall that in the fixed state case,
	  the fidelity eventually converges to one, independent of SNR.
	  We define fidelity as we did for the fixed state case in
	  Eq.~(\ref{eq:fidelity}).
	  %%%$F(\tau_{\rm{f}}) = 1-\int_{-\infty}^{\nu_{\rm{th}}}P_+(s')ds' -
	  %%%\int_{\nu_{\rm{th}}}^{\infty}P_-(s')ds'$.
	  The difference here is that because the distributions are not
	  necessarily symmetrical, the signal threshold $\nu_{\rm{th}}$ is not
	  always zero.
	  Maximizing $F$ with respect to $\nu_{\rm th}$ yields the following
	  implicit equation for $\nu_{\rm th}$
	  \begin{equation}
	  P_+(\nu_{\rm th}) = P_-(\nu_{\rm th}).
	  \end{equation}
	  As an aside we note that from this we see that we can write the
	  fidelity (optimized with respect to $\nu_{\rm th}$) in the
	  alternative form
	  \begin{equation}
	  F=\frac{1}{2}\int_{-\infty}^{+\infty}ds\,|P_+(s)-P_-(s)|
	  \end{equation}
	  Despite the complications of finite lifetime, the integrations in
	  Eq.~(\ref{eq:fidelity}) can still be carried out analytically to
	  yield
	  \begin{eqnarray}
	  F&=& \frac{1}{2}e^{\sigma^2/8}e^{-(\nu_{\rm th}+\tau)/2}\Big{\{}{\rm
		  erf}\left( \frac{\sigma^2-2(\nu_{\rm
					  th}-\tau)}{2\sqrt{2}\sigma}\right)\nl-{\rm erf}\left(
					  \frac{\sigma^2-2(\nu_{\rm th}+\tau)}{2\sqrt{2}\sigma}\right)
	\Big{\}}
	\end{eqnarray}
	The fidelity is maximized by numerically solving for $\nu_{\rm{th}}$
	such that $P_-(\nu_{\rm{th}}) = P_+(\nu_{\rm{th}})$. Using the
	correct value of $\nu_{\rm{th}}(\tau_{\rm{f}})$, it is
	straightforward to compute $F(\tau_{\rm{f}})$ and then vary
	$\tau_{\rm{f}}$ to obtain the optimal value of the integration
	time.

	\begin{figure}
	\includegraphics*[width=0.45\textwidth]{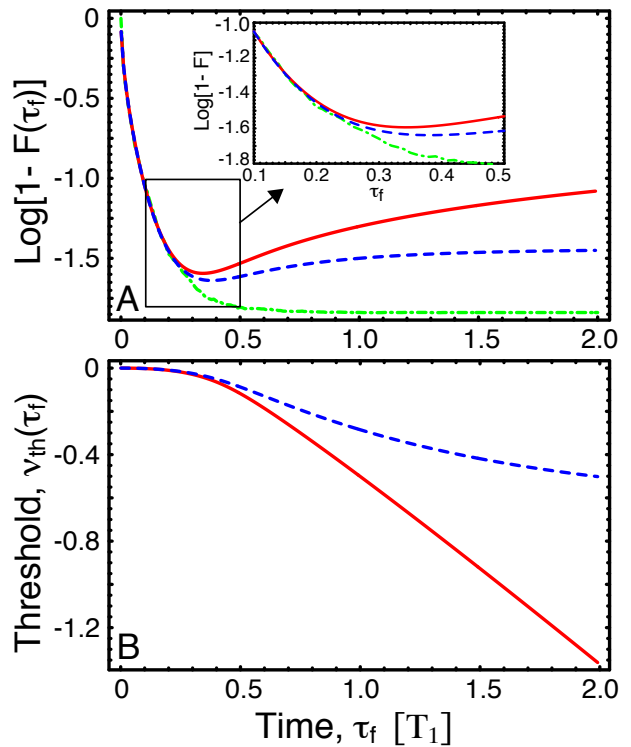}
	\caption{\label{fig:timefidelity}{\small (Color online){ A) The fidelity $F$  as a function of time
		$\tau_{\rm f}$ for all measurement protocols. The box car filtered integrated signal is the red
			(solid) line and the exponential filtered integrated signal is the
			blue (dashed) line. Here we see that both these schemes have an optimal measurement
			time and that the latter case is less sensitive to the measurement time.
			The non-linear filter is shown by the green (dashed-dotted) line.
			Here we see that it is clearly better than the other cases and that there is no optimal measurement time.
			B) The threshold $\nu_{\rm th}$ as a function of time $\tau_{\rm f}$ for both the box car filtered
			integrated signal (red solid line) and the exponential filtered integrated signal (blue dashed line).
			In both subplots the SNR after time $\tau_{\rm f} = 1$ is 10 and time is measured in units of  $T_1$. }}
	}
\end{figure}

Roughly speaking, fidelity is a measure of how separate the probability distributions are, ranging between 0 for
complete overlap and 1 for no overlap. Fidelity is limited by detector noise for short times and spontaneous
relaxation for long times. For $\tau_{\rm{f}}\ll1$, the probability of a relaxation during the measurement is
very low, and the fidelity behaves similarly to the fixed state case: it increases with increasing
$\tau_{\rm{f}}$. For longer $\tau_{\rm{f}}\approx 1$, the qubit is more and more likely to have relaxed
during the measurement and the mean of $P_+(s)$ will stop increasing linearly and in the long time limit will
actually decrease in time as can be inferred from the decrease in optimal threshold value plotted
in Fig.~\ref{fig:timefidelity} B) (red solid line). This implies the
existence of some intermediate time $\tau_{\rm{opt}}(\rm{SNR})$ that
maximizes the fidelity, this is clearly seen in the red solid line
of  Fig.~\ref{fig:timefidelity} A) where the fidelity for a SNR$=10$ has a
maximum of $0.79$ at $\tau_{\rm opt} = 0.34$.

To find this time for a given value of SNR, we compute
$F(\tau_{\rm{f}})$, then numerically solve
$dF(\tau_{\rm{f}})/d\tau_{\rm{f}} = 0$ for $\tau_{\rm
	f}=\tau_{\rm{opt}}$. The solution, $F_{\rm{opt}} \equiv
	F(\tau_{\rm{opt}})$ is the best possible fidelity
	for the simple linear box car filter; no improvement can be made from
	this value without improving the measurement apparatus or the qubit
	lifetime. This optimal fidelity is only achievable by correctly
	setting $\tau_{\rm{f}}=\tau_{\rm{opt}}$ and $\nu_{\rm{th}}$. Any
	variation of these parameters will reduce the fidelity. It should be
pointed out that while the condition $P_-(\nu_{\rm{th}}) =
P_+(\nu_{\rm{th}})$ does maximize the fidelity, it implies that the
measurement protocol is biased towards the ground state. That is,
if the qubit is prepared in the ground state we are more likely to
assign it correctly than if it was prepared in the excited state. An
unbiased measurement protocol is obtained by setting $\nu_{\rm th}$
such that $\int_{-\infty}^{\nu_{\rm th}}ds~P_-(s)  =
\int^{\infty}_{\nu_{\rm th}}ds ~P_+(s)$. Doing this results in a
slightly lower optimal fidelity.

\begin{figure}
  \includegraphics*[width=0.45\textwidth]{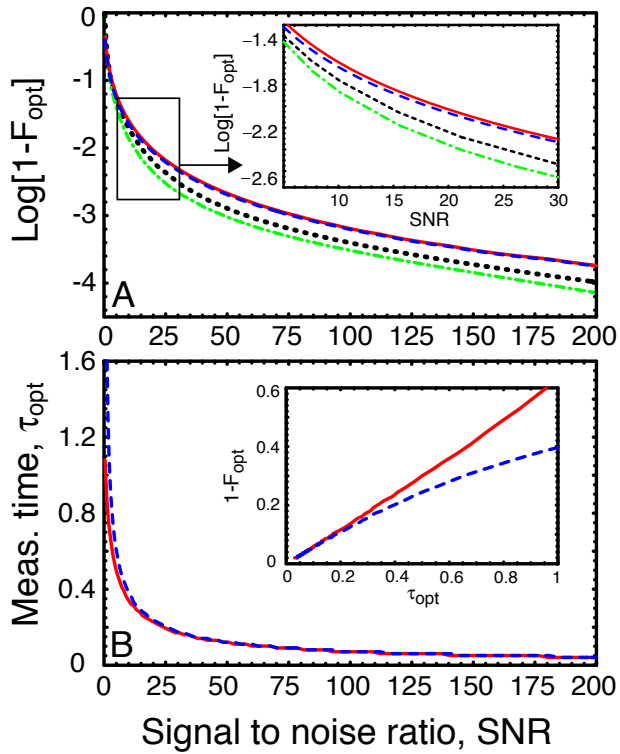}
    \caption{\label{fig:SNRfidelity}{\small (Color online) { A) Optimal fidelity $F_{\rm{opt}}$ and B)
    measurement time $\tau_{\rm{opt}}$ 
    as functions of SNR. The inset in plot B shows the optimal fidelity as a function of the measurement time.
	The red (solid) line is for the box car filter,
    blue (dashed) is for the exponential filter, black (dotted) is for the optimal linear filter,
    and the green (dashed-dotted)
    is for the non-linear filter. Here we see that by using the optimal non-linear filter
    we get an improved fidelity in comparison to the simple linear filters.
    Particular values are shown in Table~\ref{tab:fidelity}. }}
    }
\end{figure}

Since $F_{\rm{opt}}$ depends only on SNR, it is possible to first derive the required signal to noise ratio
after one lifetime, and then choose the correct measurement time and signal threshold required in order to
attain any arbitrary fidelity as shown in
 Fig.~\ref{fig:SNRfidelity} (red solid line in part A) along with the optimal measurement time
 (red solid line in part B). For example, one standard initial goal
is a fidelity of 90\%, sufficient to violate Bell's
inequalities~\cite{jennewein02}. As shown in table
\ref{tab:fidelity}, this fidelity requires a minimum SNR of 30 after
time $T_1$.

\begin{table*}
    \begin{ruledtabular}
    \begin{tabular}{lllllll}
    $F$  &$\rm{SNR}_{\rm{fixed}}$ &${\rm SNR}_{\rm BC}~(\tau_{\rm{opt}})$
    &${\rm SNR}_{\rm exp}~(\tau_{\rm{opt}})$ &${\rm SNR}_{\rm OL}$ &${\rm SNR}_{\rm NL}$ &$\tau_{\rm{arm}}$\\
    \hline
    50\% &0.50                             &1.47 (0.82)  &1.23 (1.59) &1.17        &1.1       &0.7\\
    67\% &1.0                             &4.05 (0.55)  &3.58 (0.75)  &3.10       &2.9      &0.40\\
    90\% &2.7                             &29.9 (0.17)   &28.7 (0.18)  &22.1          &18.     &0.11\\
    95\% &3.8                             &77.3 (0.087)   &75.7 (0.090)  &55.9      &48.          &0.05\\
    99\% &6.7                             &574. (0.018)  & 572. (0.018)  &420        & 269.       &0.01\\
    \end{tabular}
    \end{ruledtabular}
    \caption{\label{tab:fidelity}{\small {The required minimum signal to
    noise ratio $\rm{SNR}$ after $T_1$ and the optimal optimal measurement time, $\tau_{\rm opt}$
    for a continuous measurement of a qubit with infinite lifetime (${\rm SNR}_{\rm fixed}$),
    a simple box car linear filter (${\rm SNR}_{\rm BC}$), an exponentially
    decaying linear filter (${\rm SNR}_{\rm exp}$), the optimal linear filter (${\rm SNR}_{\rm OL}$),
    and a non-linear Bayesian filter (${\rm SNR}_{\rm NL}$).
    The last column is the maximum allowed
    waiting time $\tau_{\rm{arm}}$ for an idealized latching (instantaneous)
    measurement with infinite signal to noise
    performed after time $\tau_{\rm{arm}}$ in order to achieve the desired fidelity $F$.}}}
\end{table*}

The following argument shows how rapidly the required SNR diverges
for very high values of fidelity. %To achieve a fidelity $F=1-\eta$
%with $\eta\ll 1$, we must have sufficient SNR to be fooled only by
%those extremely early decays
%which occur in time $\tau^* \sim \eta$.
%This set of decays occurs with probability $\sim\eta$ and is the
%main source of the infidelity.
For large SNR, it is a good
approximation to set the threshold $\nu_{\rm th}=0$ and then the fidelity is approximately
\begin{equation}
F_0 = \exp(-\tau_{\rm f}/2)\erf{\sqrt{{\rm SNR} \tau_{\rm f}
/2}} \label{eq:F0}
\end{equation}
%which is consistent with the intuitive picture just described.
Optimizing this with respect to $\tau_{\rm f}$ and then using the
asymptotic form for the error function of large argument yields the
following expression for the optimal integration time
\begin{equation}
\tau_{\rm opt} \approx \frac{2}{\rm SNR}\left\{x_0 -
\frac{1}{2}\ln(x_0) \right\}
\end{equation}
where $x_0\equiv \ln\left(\frac{1+{\rm SNR}}{\sqrt{\pi}}  \right)$.

Approximating the error function in Eq.~(\ref{eq:F0}) by unity and
neglecting $\ln(x_0)$ relative to $x_0$ leads to the following
simple asymptotic form for large SNR
\begin{equation}
F_0\sim 1-{\frac{1}{\rm SNR}}\ln\left( \frac{1+{\rm SNR}}{\sqrt{\pi}}
\right).
\end{equation}
This can be rewritten as in terms of the optimal measurement time as $F_0\sim 1-\tau_{\rm opt}/2$. Here we see that
to achieve the optimal, the measurement
must be completed in a time that decreases linearly with the desired fidelity.
The slight deviations form this zeroth order result is shown in the inset of Fig. \ref{fig:SNRfidelity} B. This result is
consistent with the intuitive picture that to achieve a fidelity $F=1-\eta$ with $\eta\ll 1$, we must have
sufficient SNR to be fooled only by those extremely early decays which occur in time $\tau^* \sim 2\eta$. This set of decays occurs with probability $\sim2\eta$ and is the main source of the infidelity.

\section{\label{sec:Exp}Exponential decaying linear filter}

In this section we consider the case when we include an exponential
weighting factor in our integrated signal $s$. This is chosen as this would be the optimal linear
filter if our signal was simply a decaying exponential with a random initial amplitude, this can
be proven by minimizing our estimate of the amplitude in a least square sense.
However, even though our signal on average is of this form, in one particular run it is not. Thus
this will not be the optimal linear filter for estimating the initial state of the qubit. For this
reason we introduce an integration time $\tau_{\rm f}$ and optimize the
measurement fidelity over this. That is, the signal $s$ for this filter is given by,
\begin{equation}
s = \int_0^{\tau_{\rm f}} d\tau~ \psi(\tau) e ^{-\tau}.
\end{equation} As in the last section we first determine a possible $s$
given that the qubit started in the
excited state. Doing this gives
\begin{eqnarray}
s_\pm &= & \int_0^{\tau_{\rm f}} d\tau\Big{[} \theta(\tau_{\rm
d} -\tau) -  \theta(\tau - \tau_{\rm d}) \Big{]}e
^{-\tau}\nl+ \sqrt{{\rm SNR}^{-1}}\int_0^{\tau_{\rm f}} dW(\tau)
~e ^{- \tau}
\end{eqnarray} and by treating the noise integral as simply a linear combination of infinitesimal Gaussian variables gives
\begin{eqnarray}
s_\pm &=&\Big{[}1 -2 e^{-\tau_{\rm d}} + e^{-\tau_{\rm f}}
\Big{]}\theta(\tau_{\rm f} - \tau_{\rm d})
\nl+a~\theta(\tau_{\rm d}-\tau_{\rm f})+  X_\mathrm{G}[0,
\sigma^2],
\end{eqnarray}
where $\sigma = \sqrt{ (1-e^{-2\tau_{\rm f}})/2~{\rm SNR}}$ and $a
= 1-e^{-\tau_{\rm f}}$. Here we see that unlike before as the
measurement time becomes large the variance saturates at $(2 ~{\rm
SNR})^{-1}$ rather then continuing to increase linearly with time.
That is,  we have designed our filter such that when all the
information about the qubit state has been lost into the $T_1$
environment the noise in the integrated signal will remain constant.

 Following the same procedure as before the excited state distribution is
\begin{eqnarray}
P_{+}(s) & =& \frac{1}{ 4  } { \left[
\text{erf}\left(\frac{{a+s}}{\sqrt{2} \sigma}\right)+
   \text{erf}\left(\frac{{a-s}}{\sqrt{2} \sigma}\right)\right]}\nl+\frac{1}{\sqrt{2 \pi } \sigma} \exp\left[{{-\frac{  \left(s-a\right)^2}{2\sigma^2}-\tau_{\rm f} }}\right].
\end{eqnarray}
Repeating the above for the ground state initial condition gives
$s_f =-a +  X_\mathrm{G}[0, \sigma^2]$ and a  ground state distribution
of the form
\begin{eqnarray}
P_{-}(s)  = \frac{1}{\sqrt{2 \pi } \sigma} {\exp\left[{-\frac{
\left(s+a\right)^2}{2\sigma^2}}\right]}.
\end{eqnarray}
With the above two distributions the fidelity, as before, can be
solved analytically and in terms of the $\nu_{\rm th}$ is
\begin{widetext}
\begin{eqnarray}
F = \frac{\sigma }{  \sqrt{8\pi }}\left(\exp\left[-\frac{(a-\nu_{\rm
th} )^2}{2\sigma^2
   }\right]-\exp\left[{-\frac{(a+\nu_{\rm th}  )^2}{2 \sigma^2}}\right]\right)+\frac{(e^{-\tau_{\rm f}} +1-\nu_{\rm th} )}{4 }
   \left[\text{erf}\left(\frac{a-\nu_{\rm th} }{\sqrt{2}\sigma}\right)+\text{erf}\left(  \frac{a+\nu_{\rm th}}{\sqrt{2}\sigma }\right)\right].
   \end{eqnarray}\end{widetext}

Using the same procedure as before we can numerically determine the $\nu_{\rm th}$ which maximizes the fidelity. For
a SNR of 10 (in time $T_1$) the fidelity as a function of measurement time is shown in Fig.
\ref{fig:timefidelity} A) as a blue dashed line. Here we see as before there is an optimal measurement time. To
measure any longer than this time results in a lower fidelity. The optimal fidelity and measurement time are
shown in Fig. \ref{fig:SNRfidelity} (blue dashed line)  as a function of the SNR. Here we see that by using this
filter, the fidelity is slightly better than the simplest case (see Table~\ref{tab:fidelity} for some values),
but more importantly the curvature of the fidelity at $\tau_{\rm opt}$ is less. This means that this filter is
less sensitive to errors in the measurement time. That is, this protocol would be more practical  to implement
than the simple box car integrated signal of Sec. \ref{sec:Flat}.

\section{\label{sec:optL} Optimal linear filter}

In this section we calculate the optimal linear filter for estimating the initial state of the qubit.
We define the linear signal $s$ by the relation
\begin{equation}
s = \int_0^{\infty} k(\tau) \psi(\tau) d\tau,
\end{equation} where the kernel $k(t)$ is unknown and is determined by maximizing the measurement
fidelity. This, as before, is defined as the difference between the probability of us making a correct
assignment and an incorrect assignment [\eqrf{eq:fidelity}]. The assignment criteria we use is again
if $s$ is above $\nu_{\rm th}$
then we say the qubit was initially up and if it is below $\nu_{\rm th}$ then it was down. $\nu_{\rm th}$ like $k(t)$ is
determined by the maximization procedure which we will describe now.

\begin{figure}
     \includegraphics*[width=0.45\textwidth]{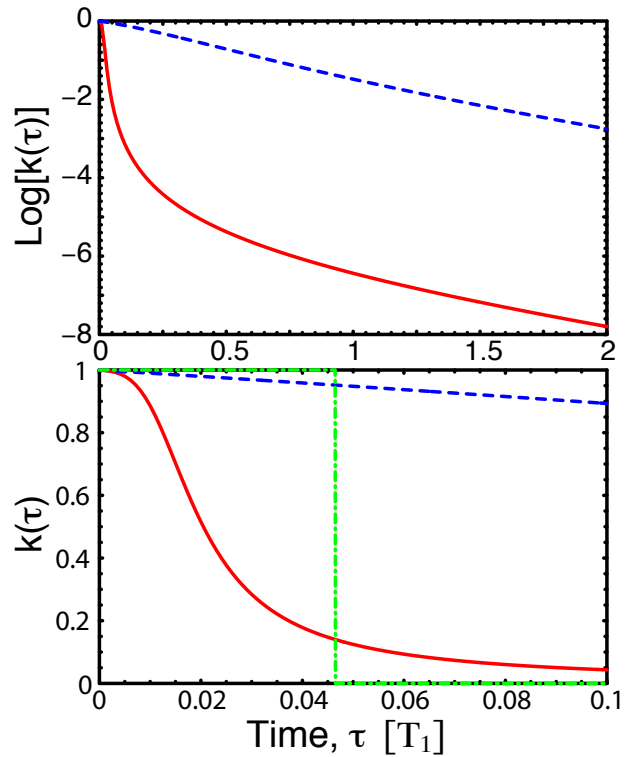}
    \caption{\label{fig:kopt}(Color online) {\small The optimal linear filter for a SNR of $1.0$ blue (dashed) and SNR of
    $172$ red (solid) as a function of time. The green (dashed-dotted) line in part B) is
    the optimal box car linear filter for a SNR of 172 ($\tau_{\rm opt} = 0.046$).
    }}
\end{figure}

For an unknown kernel the signal conditioned on the qubit being initially up will be given by
\begin{equation}
s_{+} = 2 a(\tau_{\rm d}) - a(\infty) +  X_\mathrm{G}[0, \sigma^2],
\end{equation}
where
\begin{eqnarray}
a(l) &=& \int_0^l k(\tau') d\tau' \\
\sigma &=& \sqrt{\int_0^\infty k^2(\tau') d\tau'/{\rm SNR}}. \label{optsigma}
\end{eqnarray}
If the qubit was initially in the ground state then the signal would be
\begin{equation}
s_{-} = - a(\infty) +  X_\mathrm{G}[0,\sigma^2].
\end{equation}
>From the above equations the excited state and ground state probability distribution for the signal
$s$ are
\begin{eqnarray}
P_+(s) &=& \int_0^\infty  \frac{
\exp[-\tau_{\rm d}-(s-2a(\tau_{\rm d})+a(\infty) )^2/2\sigma^2]}{\sqrt{2 \pi \sigma^2}}d\tau_{\rm d}, \nl\\
P_-(s) &=& \frac{
\exp[-(s+a(\infty) )^2/2\sigma^2]}{\sqrt{2 \pi \sigma^2}}
\end{eqnarray}
and by using \eqrf{eq:fidelity} the fidelity is
\begin{eqnarray} \label{optfid}
F  &=& \frac{1}{2} \int_0^\infty e^{-\tau_{\rm d}} {\rm erf}\Big{(} \frac{2 a(\tau_{\rm d}) -a(\infty) -\nu_{\rm th}}{\sqrt{2} \sigma}\Big{)}d\tau_{\rm d}, \nl
- \frac{1}{2}~ {\rm erf} \Big{(} \frac{-a(\infty) -\nu_{\rm th}}{\sqrt{2} \sigma}\big{)}.
\end{eqnarray} Maximizing this gives the following set of coupled differential equations
\begin{eqnarray}
d_\tau a (\tau) &=& k(\tau), \\
d_\tau k(\tau) &=&  -\exp\{-\tau-2a(\tau) [a(\tau) - a(\infty) -\nu_{\rm th}]/\sigma^2\},\nl
\end{eqnarray}
with the initial conditions $k(0) = 1$, $a(0)=0$, and the boundary condition $k(\infty) = 0$. It is this latter
condition which determines $\nu_{\rm th}$. This system of equations
can be solved numerically using a shooting method \cite{NR}. The results are shown in Fig. \ref{fig:kopt} for a
{\rm SNR} of $1.0$ and $172$. Here we see that for the small ${\rm SNR}$, the kernel $k(\tau)$
can be approximated well by
$\exp(-\beta \tau)$ where $\beta$  is a fit parameter that is approximately equal to $1+{\rm SNR}/2$.
In the large ${\rm SNR}$ limit the $k(\tau)$ cannot be fit by an exponential.
For illustrative purposes the optimal linear filter is compared with the optimal box car linear filter in
Fig \ref{fig:kopt} B) for a SNR of 172. Here we see that the time when the box car linear filter turns off,
is comparable with the time scale of the optimal linear filter.

Numerically solving for $k(\tau)$ for a given $\sigma$ we can use \eqrfs{optsigma}{optfid} to plot the
fidelity of the optimal
linear filter as a function of the SNR. This is shown in Fig. \ref{fig:SNRfidelity} as a dotted black line and in Table
\ref{tab:fidelity} as column 4. Here we see that
the optimal linear filter out performs the other linear filters and is almost as good as the optimal filter
(which is non-linear) and is described in the next section.

\section{\label{sec:NL}Optimal non-linear filter}

In the previous sections we considered only the case where one looks at or only has access to the
integrated signal, $s(\tau)$.  Here we assume that we have access to the full record $\Psi = \{\psi(\tau)\}$  and ask how
much better can we do with a non-linear filter. In particular, given this record
what is our best guess at the initial state of the qubit. Mathematically, our best guess can
be represented by the probability distribution $ P(i_0| \Psi) $. This is the  probability that
the initial condition  $i_0$ is  $\pm 1$ given the record, $\Psi$. As with all probability
distributions, this will range from 0 to 1, and the closer it is to one the more we are certain that
the qubit was in the initial state $i_0 =\pm1$.
This by definition is the optimal protocol as it is the best estimate of $i_0$ given the complete set of
information available. To find this distribution we use Bayes theorem,
\begin{equation}\label{Bayes}
 P(i_0| \Psi) = \frac{P( \Psi|i_0) P(i_0)}{\sum_{i_0} P( \Psi|i_0) P(i_0)},
 \end{equation}  where $P(i_0)$ is the initial probability distribution and
 $P( \Psi|i_0)$ is the probability that we would measure record $\Psi$ given that the experiment
 was initially prepared in state $i_0$. We will assume that our experiment can prepare unbiased
 initial states and as such we take  $P(i_0\pm1)=1/2$.  The conditional distribution $P( \Psi|i_0)$
 is the probability that we would measure record, $\Psi$, given that the experiment was initially prepared in state $i_0$.
 Thus to calculate  $ P(i_0| \Psi)$  all we need to do is calculate  $ P( \Psi| i_0)$.
 This is the non-trivial step, but can be determined by  \begin{equation}
P( \Psi|i_0) = \sum_{p} P(\Psi|i_p)P(i_p| i_0),
\end{equation}
where $p$ labels a possible realization of the qubit
trajectory. That is, $P( \Psi|i_0)$ can be determined by taking the ensemble average of
$P(\Psi|i_p)$ for all possible realizations,  $i_p$ ( $i_p$ is given by \eqrf{if_excited}  if the qubit is
initially in the excited state or $i_p=-1$ if initially in the ground state).
>From our simple model for the noise,  \eqrf{psigz},  $P(\Psi|i_p)$ is simply a multiplication of
many Gaussians each centered on the instantaneous value of $i_p$.

For an excited state initial condition, the probability of getting record $\Psi$ will be given by
\begin{widetext}
 \begin{eqnarray}
 P( \Psi|i_0 =1)& =& A \int_{0}^{\tau_{\rm f}}  d \tau_{\rm d}~ e^{-\tau_{\rm d}}
\exp\left[-\int_0^{\tau_{\rm d}} d\tau(\psi(\tau) - 1 )^2{\rm
SNR} /2 \right] \exp\left[-\int_{\tau_{\rm d}}^{\tau_{\rm f}}
d\tau(\psi(\tau) + 1 )^2 {\rm SNR}/2 \right] \nl+ e^{-\tau_{\rm
f}} \exp\left[-\int_0^{\tau_{\rm f}} d\tau(\psi(\tau) - 1 )^2
{\rm SNR}/2 \right] .
\end{eqnarray}  Here the first term represents all possible trajectories that have had a decay in
time $\tau_{\rm f}$ (the integration time), the second term represents all the trajectories that did not
decay in this time, and $A$ is the normalization constant.
If the qubit were  initially in the ground state then the  probability of getting record $\Psi$ will be given by
 \begin{eqnarray}
 P( \Psi|i_0 = -1)& =& A
\exp\left[-\int_0^{\tau_{\rm f}}d\tau  (\psi(\tau) + 1 )^2{\rm
SNR}/2 \right ] .
\end{eqnarray}

Now that we have expressions for  $P( \Psi|i_0)$  all that we need to do to get our best estimate of the
initial state is to use \eqrf {Bayes}. Doing this gives
 \begin{eqnarray} \label{OPUP}
 P( i_0 =1|\Psi ) &=& \frac{1}{\rm Norm} \left[\int_{0}^{\tau_{\rm f}} d \tau_{\rm d} ~ e^{-\tau_{\rm d}}
\exp\{ [s(\tau_{\rm d}, 0)-s(\tau,\tau_{\rm d})] {\rm SNR}\}
+ e^{-\tau_{\rm f}} \exp[s(\tau_{\rm f},0) {\rm SNR} ]\right]
\nl \\\label{OPDOWN}
 P( i_0 = - 1|\Psi ) &= &\frac{1}{\rm Norm}
\exp[-s(\tau_{\rm f},0) {\rm SNR} ]
\end{eqnarray}  where the two time integrated signal $s(\tau, \tau')$ is
\begin{equation}\label{stt}
s(\tau, \tau') =  \int_{\tau'}^\tau d\tau''~\psi(\tau'')
\end{equation}
and the norm is simply
 \begin{equation}
{\rm Norm} = \int_{0}^{\tau_{\rm f}}  d \tau_{\rm d}~
e^{-\tau_{\rm d}} \exp[ (s(\tau_{\rm d}, 0)
-s(\tau,\tau_{\rm d})) {\rm SNR} ] + e^{-\tau_{\rm f}}
\exp[s(\tau_{\rm f},0) {\rm SNR} ]+ \exp[-s(\tau_{\rm f},0) {\rm
SNR} ],
\end{equation} \end{widetext}

Here we see that to solve this equation we need to evaluate a double integral over a stochastic process.
This is impractical to solve numerically, however, as shown in the Appendix we can easily recast
these integrals in terms of two sets of two coupled stochastic differential equations which require
similar computational resources to that used with the linear filters.

\begin{figure}
  \includegraphics*[width=0.45\textwidth]{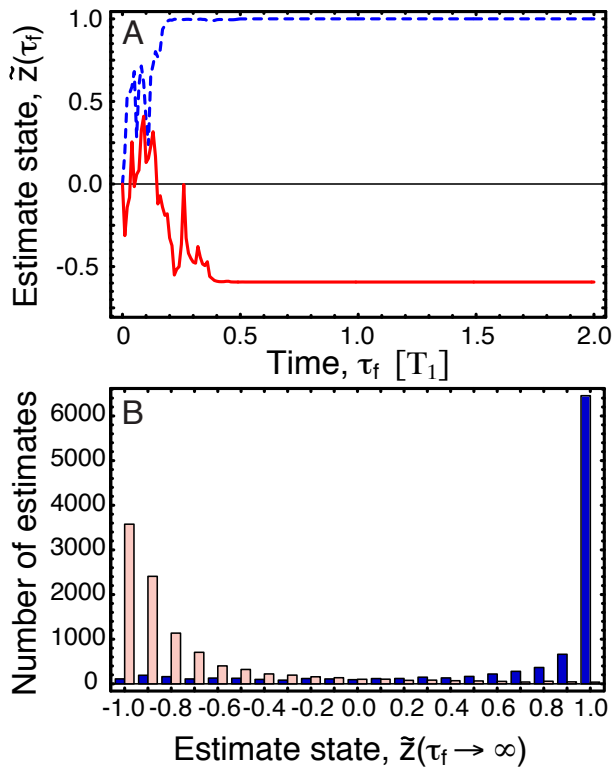}
    \caption{\label{fig:zest}{\small { A) A typical trajectory for the estimated initial condition,
    $\tilde{z} (\tau_{\rm f})$, given a record $\Psi$ which is randomly generated for an excited
    state initial condition (blue dashed line) and a ground  state initial condition (red solid line).
    B) A histogram of $\tilde{z} (\tau_{\rm f} \rightarrow \infty)$ for $10^4$ trajectories for both
    an excited state initial condition (dark blue bars column 1) and a ground state initial condition (light red bars column 2).
    The SNR after $\tau_{\rm f} =1$ is 10 and time is measured in units of $T_1$.}
    }}
\end{figure}

 To show a
typical trajectory for this estimated initial condition, we randomly generated  records for both an excited and
ground state initial condition.  Rather then plotting both $P( i_0 =1|\Psi )$ and $P( i_0 =-1|\Psi )$
we define  $\tilde{z} =
P( i_0 =1|\Psi )-P( i_0 =-1|\Psi ) $ (this is the estimator that replaces $s$ used in the linear filters),
this will range from $-1$ to $1$ and the closer it is to one of these
limits, the more certain we are that the initial condition which generated $\Psi$ is this value.
The results of this simulation are shown in Fig.
\ref{fig:zest} A). The red (solid) line corresponds to the case when the initial state was the ground state and
the blue (dashed) line is for the excited state.
Here we see that  for this typical trajectory our estimate is
fairly good at predicting the real initial condition and we are almost certain for the excited case. However
this is only a typical trajectory and to get a better understanding of the predictability of this method we use
the same fidelity measure as before,  that is we subtract our wrong guesses from our correct guesses. To be more
specific, we define the following assignment procedure: if $\tilde{z}> 0$ then we assign the qubit as up and if
$\tilde{z}< 0$ we assign it as down. If $\tilde{z}$ is equal to zero we ignore the result (or flip an unbiased
coin to make our decision). Given this assignment criteria we can define the fidelity as
\begin{equation}
F = \lim_{M\rightarrow \infty}
\frac{1}{2M}\Big{[}\sum_{\tilde{z}_{\e}> 0}-\sum_{\tilde{z}_{\e}< 0}
+\sum_{\tilde{z}_{\g}< 0}-\sum_{\tilde{z}_{\g}> 0}\Big{]}
\end{equation} where $M$ is the number of randomly generated $\Psi$ for
both $\e$ and $\g$ initial conditions.  This is shown in Figs. \ref{fig:timefidelity} A)
and \ref{fig:SNRfidelity} A) as the green
(dashed-dotted) line for $M =10^4$ (specific optimal fidelities are listed in Table~\ref{tab:fidelity}).
Here we see that the fidelity is
always better than the other cases
%(the same optimal fidelity can be achieved at approximately half the SNR)
and that there is no optimal measurement time. That is, unlike the box car filter and the exponential
linear filter, the fidelity is a
non-decreasing function of the integration time.

  It should be noted that as in the previous protocols, this is a biased measurement which
  favors ground state preparations. This can be seen by looking closely at Fig. \ref{fig:zest} B).
  This figure shows a histogram of $10^4$ estimates when the system is prepared in the excited state
  (dark blue bars, column 1) and the ground state (light red bars, column 2) for a $ {\rm SNR =10}$.  If we use our
  assignment procedure and subtract the wrong guesses from the correct guesses for each prepared initial
  state we get a fidelity of 0.92 for the ground state and 0.76 for the excited state (an average fidelity of 0.84).
  This is because when the qubit is prepared in the excited state, rare, early decays will fool the detector.

\section{\label{sec:latch}Comparison with latching measurement}

The challenges from qubit relaxation to attaining high fidelity in a
continuous measurement also exist in other measurement schemes. For
comparison, consider a latching measurement where the qubit state triggers a classical switching event in the
measurement apparatus \cite{martinis02,martinis_katz06,siddiqi04,siddiqi05,siddiqi06}.  Such a measurement has
the advantage that the detector state stays latched for a very long
time, so that noise from subsequent amplification stages is
completely negligible and the SNR is effectively infinite.  We can
roughly model this process as an instantaneous measurement with no
errors but a finite arming time $t_{\rm{arm}}$ needed to set up the
pre-latched state of the detector.  We assume that the arming stage
occurs after the qubit is prepared but before it is measured. The
measurement of a qubit in the excited state will be wrong if the
qubit relaxes before the measurement is made. The probability that
the qubit relaxes during $t_{\rm{arm}}$ rises exponentially towards
1, so the fidelity falls exponentially,
$F(\tau_{\rm{arm}})=e^{-\tau_{\rm{arm}}}$. As shown in Table
\ref{tab:fidelity}, $\tau_{\rm{arm}}$ must be slightly smaller than
$\tau_{\rm{opt}}$. Although these times are different in that
$\tau_{\rm{arm}}$ is a maximum value whereas $\tau_{\rm{opt}}$ is an
optimal value for a given SNR, they provide a valuable comparison
between the measurement schemes.

Of course, no latching measurement is truly instantaneous with
perfect accuracy in translating the qubit state into latching events.
Moreover, a latching measurement usually induces some mixing of states, so the actual fidelity may be lower than
for an equivalent continuous QND measurement. Both measurement schemes have their own strengths and weaknesses,
but in either case, qubit decay can be a significant limiting factor on the fidelity.

\section{\label{sec:conclusion}Conclusion}

We have examined the effect of qubit relaxation on the estimation of the qubit
initial state (excited or ground) using a continuous-in-time noisy quantum non-demolition  measurement for four different
measurement protocols. In these protocols the measurement results are integrated with a box car linear filter, an exponentially decaying
filter, an optimal linear filter  and a
non-linear Bayesian filter which by definition is the optimal theoretical filter. We found that in all these
protocols there exists a theoretical limit on the measurement fidelity. The determining factor of this limit is
the signal to noise ratio of the measurement. Our results are summarized in Table~\ref{tab:fidelity} where we
see that the non-linear filter reaches the same fidelity as the linear filters even for substantially lower required signal to noise ratio.
Lastly we compare the continuous quantum non-demolition  results with latching measurements and found that  there is a
quantitatively different but qualitatively similar limit on the fidelity of latching measurements also due to
relaxation. The signal to noise ratio required to do successful qubit single shot quantum non-demolition  measurements
should be attainable in the near future, and there is no fundamental reason why significantly higher fidelity
measurements cannot be performed.

\begin{acknowledgments}
We thank Michel Devoret, Isaac Chuang, Alexandre Blais, Jens Koch, Andrew Houck, and David Schuster for
discussions. This work was supported in part by NSA under ARO Contract No W911NF-05-1-0365, and the NSF
under grants ITR-0325580, DMR-0342157 and DMR-0603369, and the W. M. Keck Foundation.
\end{acknowledgments}

\bigskip

\appendix

\section{Numerical procedure used by the optimal non-linear filter }  \label{optimal}

In this appendix we present the method used to simulate the optimal
non-linear filter. This filter requires simulating
\eqrfs{OPUP}{OPDOWN} which contains a double integral over a
stochastic process $\psi(\tau)$.  This is not practical numerically
and a much better method can be implemented by deriving a set of
coupled stochastic differential equations. To do this we start by
rewriting \eqrfs{OPUP}{OPDOWN}  as
 \begin{eqnarray} \label{OPUP2}
 P( i_0 =1|\Psi ) &=& \frac{\bar{P}_{+1,\psi}}{\bar{P}_{-1,\psi} + \bar{P}_{+1,\psi}}, \\\label{OPDOWN2}
 P( i_0 = - 1|\Psi ) &= &\frac{ \bar{P}_{-1,\psi} }{\bar{P}_{-1,\psi} + \bar{P}_{+1,\psi} },
\end{eqnarray}  where \begin{widetext}
 \begin{eqnarray} \label{OPUPbar}
\bar{P}_{+1,\psi} &=&  e^{-\tau_{\rm f}} \exp[s(\tau_{\rm f},0)
{\rm SNR} ] +\int_{0}^{\tau_{\rm f}}  d \tau_{\rm
d}~e^{-\tau_{\rm d}} \exp\{ [s(\tau_{\rm d},
0)-s(\tau_{\rm f},\tau_{\rm d})] {\rm SNR}\}, \\\label{OPDOWNbar}
\bar{P}_{-1,\psi}&= & \exp[-s(\tau_{\rm f},0) {\rm SNR} ],
\end{eqnarray}
and the signal $s(\tau, \tau')$ is defined in \eqrf{stt}. We first
differentiate $\bar{P}_{+1,\psi}$ with respect to $\tau_{\rm f}$
and by defining
 \begin{eqnarray}
\lambda_{+1,\psi} &=&  e^{-\tau_{\rm f}} \exp[s(\tau_{\rm f},0)
{\rm SNR} ] -\int_{0}^{\tau_{\rm f}} d \tau_{\rm
d}~e^{-\tau_{\rm d}}\exp\{ [s(\tau_{\rm d},
0)-s(\tau_{\rm f},\tau_{\rm d})] {\rm SNR}\},\end{eqnarray}
\end{widetext} we get the following set of coupled  stochastic
differential equations
 \begin{eqnarray} \label{optset1}
\frac{d}{d\tau_{\rm f}}\bar{P}_{+1,\psi} &=&  {\rm SNR}~ \psi(\tau)
\lambda_{+1,\psi},\\\label{optset2} \frac{d}{d\tau_{\rm f}}\lambda_{+1,\psi}&= & {\rm
SNR}~ \psi(\tau) \bar{P}_{+1,\psi} - ( \lambda_{+1,\psi}+
\bar{P}_{+1,\psi}), \hspace{1cm}\end{eqnarray} and the initial conditions
$\lambda_{+1,\psi} = 1$ and  $\bar{P}_{+1,\psi} = 1$.  If we use
\eqrf{OPDOWNbar}  the two additional coupled equations are
\begin{eqnarray} \label{optset1a}
\frac{d}{d\tau_{\rm f}}\bar{P}_{-1,\psi} &=&  {\rm SNR}~ \psi(\tau)
\lambda_{-1,\psi},\\\label{optset2a} \frac{d}{d\tau_{\rm f}}\lambda_{-1,\psi}&= & {\rm
SNR}~ \psi(\tau) \bar{P}_{-1,\psi}, \end{eqnarray} with
 initial conditions $\lambda_{-1,\psi} = -1$ and
$\bar{P}_{-1,\psi} = 1$. Note that since there is no relaxation, we do not need to have two equations for the ground state,
we can just use $d_{\tau_{\rm f}}\bar{P}_{-1,\psi} = -{\rm
SNR}~ \psi(\tau) \bar{P}_{-1,\psi} $, but to keep the problem
symmetrical we have decided to leave both equations in. This makes it easier to
extend the theory to cases where upward jumps are possible.

Thus to simulate \eqrfs{OPUP}{OPDOWN}  we simply solve the above two sets of two coupled differential equations and
then combine them using \eqrfs{OPUP2}{OPDOWN2}. Note an equivalent derivation of
these equations can be made by using the Kushner-Stratonovich equation \cite{Mcg74}  and then simply using
Bayes theorem to invert these equations for estimating  unknown parameters \cite{GamWis01}.

\end{document}